\documentclass[10pt]{iopart}

\usepackage{iopams}

\usepackage{graphicx}

\usepackage{hyperref}

\def\bg#1\eg{\begin{eqnarray}#1\end{eqnarray}}

\newcommand{\I}{\rmi}

\begin{document}

\title{Controlling induced coherence for quantum imaging}

\author{Mikhail~I.~Kolobov}
\address{Univ. Lille, CNRS, UMR 8523 - PhLAM - Physique des Lasers Atomes et Mol\'ecules, F-59000 Lille, France.}

\author{Enno Giese, Samuel Lemieux, Robert Fickler}
\ead{egiese@uottawa.ca}
\address{Department of Physics, University of Ottawa, 25 Templeton Street, Ottawa, Ontario K1N 6N5, Canada.}

\author{Robert W. Boyd}
\address{Department of Physics, University of Ottawa, 25 Templeton Street, Ottawa, Ontario K1N 6N5, Canada.}
\address{Institute of Optics, University of Rochester, Rochester, New York 14627, USA.}

\begin{abstract}
Induced coherence in parametric down-conversion between two coherently pumped nonlinear crystals that share
a common idler mode can be used as an imaging technique.
Based on the interference between the two signal modes of the crystals, an image can be reconstructed.
By obtaining an expression for the interference pattern that is valid in both the low- and the high-gain regimes of parametric down-conversion, we show how the coherence of the light emitted by the two crystals can be controlled.
With our comprehensive analysis we provide deeper insight into recent discussions about the application of induced coherence to imaging in different regimes.
Moreover, we propose a scheme for optimizing the visibility of the interference pattern so that it directly corresponds to the degree of coherence of the light generated in the two crystals.
We find that this scheme leads in the high-gain regime to a visibility arbitrarily close to unity.
\end{abstract}

\vspace{2pc}

\ioptwocol

\section{Introduction}

Induced coherence without induced emission~\cite{Zou91,Wang91} is a remarkable phenomenon in which nonclassical features seem to manifest themselves in ordinary interference patterns instead of higher-order correlation functions.
The experiment consists of two nonlinear crystals that share a common idler mode and are coherently pumped.
By placing an object into the idler mode between the two crystals, its image can be obtained from the interference of the two output signal modes.

In light of the application of nonlinear interferometers to spectroscopy~\cite{Kalashnikov16}, the concept of induced coherence has been recently applied to quantum imaging~\cite{Barreto-Lemos14,Lahiri15}, whereas the original experiment~\cite{Zou91} focussed on the physical principle behind the effect.
The imaging experiments have been performed in the low-gain regime of parametric down-conversion, where a quantum description and interpretation is the only possibility.
However, induced coherence is not restricted to this regime~\cite{Belinsky92} and also persists for higher gain~\cite{Wiseman00}.
In particular, essential properties for imaging, such as the the signal-to-noise ratio, improve for the latter~\cite{Shapiro15}.

In this article we provide a comprehensive treatment of induced coherence, compare and contrast different regimes of parametric down-conversion, and outline how to optimize the properties of such a setup.

\subsection{Setup and low-gain interpretation}
\label{subsec_Setup_and_low-gain_interpretation}
The first experiments~\cite{Zou91,Wang91} of induced coherence without induced emission were performed with a small parametric gain.
In fact, only in the low-gain regime induced emission is suppressed and an intuitive quantum interpretation can be given.
In this section we recapitulate the setup used in~\cite{Zou91} and explain the results for low gain with the quantum-mechanical arguments.

\begin{figure}[htb] \centering
\includegraphics{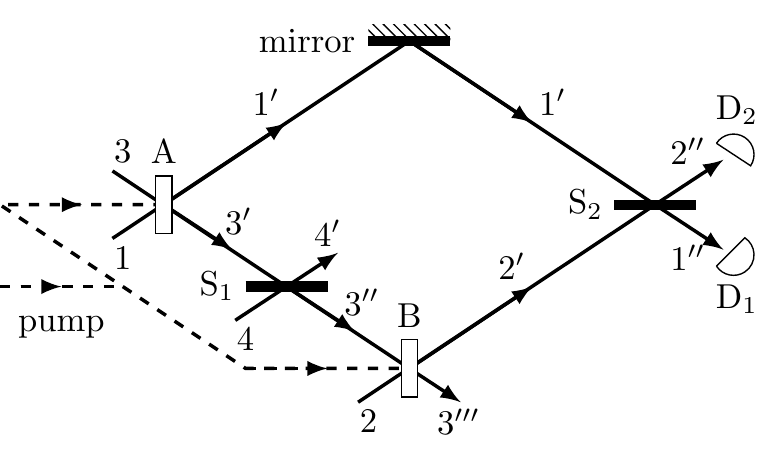}
    \caption{
    Setup of an experiment to observe induced coherence.
    Two second-order nonlinear crystals A and B are pumped by the same coherent pump laser.
    Both crystals share the same idler mode, whereas their two signal modes interfere at a beam splitter S$_2$ and are detected by two detectors D$_1$ and D$_2$.
    The coherence induced in crystal B can be controlled by introducing a filter S$_1$ into the idler mode between the two crystals.
	}
\label{fig_Setup}
\end{figure}

A simplified scheme of this experiment is shown in \fref{fig_Setup}.
It consists of two nonlinear crystals A and B optically pumped by a light wave obtained from the same laser source.
In the process of parametric down-conversion, each crystal emits its own signal and idler waves.
The two signal waves, emitted into separate modes $1'$ and $2'$, are brought to interference through the use of a 50:50 beam splitter S$_2$ and are detected at the detectors D$_1$ and D$_2$.
The two idler waves are emitted into the same spatial mode such that crystal A seeds the input idler mode of crystal B.
By inserting a filter S$_1$ with intensity transmittance $T$ into the idler mode between the two crystals, one can control the strength of the coupling caused by the idler wave propagating from A to B.
Obviously, for $T=0$ the two crystals emit independently and no coherence between the two signal modes $1'$ and $2'$ can be observed when they interfere after the beam splitter S$_2$.
However, for a nonvanishing transmittance the wave transmitted by S$_1$ seeds crystal B and therefore establishes coherence between the modes $1'$ and $2'$, which reflects itself in a nonzero visibility in the interference pattern.

The interference in~\cite{Zou91} was observed in the low-gain regime of spontaneous parametric down-conversion, in which case the emission of photons in crystals A and B seems to be completely uncorrelated.
If only a single pair of photons is detected, one can argue that the quantum state of the system is a superposition of single-photon state created in crystal A \emph{and} a single-photon state created in crystal B.
The visibility of the interference pattern in this regime depends linearly on the amplitude transmittance $\sqrt{T}$ of the filter S$_1$.
The interference phenomenon in the single-photon regime and the dependence on $T$ was explained in~\cite{Zou91,Barreto-Lemos14} in terms of indistinguishability of the photons emitted by the crystals A and B.
Indeed, for $T=1$ it is impossible to determine which of two crystals contributed the photon detected at D$_1$ or D$_2$.
In the opposite limit of $T=0$ it becomes in principle possible to determine with absolute certainty the origin of the detected photon, wiping out the interference.
This explanation is also used in more recent and complicated applications, for example in~\cite{Heuer15}.

\subsection{Connection to previous work and outline}

The experiment~\cite{Zou91} has attracted great attention after its publication~\cite{Belinsky92,Burlakov97,Korystov01,Burlakov02} and continues to influence the quantum optics community to date~\cite{Barreto-Lemos14,Shapiro15,Heuer15}.
We shall mention here only a few papers that are closely related to our analysis.

In~\cite{Belinsky92} the experiment from~\cite{Zou91} is analyzed for arbitrary parametric gains in the two crystals and a general expression for the visibility is obtained.
The authors conclude that the interference effect observed in~\cite{Zou91} in the low-gain regime will persist in the high-gain regime and pointed out that the visibility becomes diminished if all the modes are seeded.
With a very similar treatment,~\cite{Wiseman00} shows that the first-order coherence function of the two modes $1'$ and $2'$ approaches unity in the high-gain regime.

In a recent experiment~\cite{Barreto-Lemos14} the original idea from~\cite{Zou91} was generalized to a spatially multi-mode configuration that allowed them to develop an imaging technique in which the object is sensitive only to the frequency of the idler, but not the one of the signal.
The authors were able to obtain an image by detecting the signal photons, even though only the idler photons were sensitive to the object.
In~\cite{Shapiro15} it was argued that almost all features of the experiment~\cite{Barreto-Lemos14} are present if crystal A is pumped in the high-gain regime.
In fact, some of the imaging properties such as the signal-to-noise ratio is improved in this limit even though the visibility of the interference pattern is decreased.

In \sref{sec_Interference_signal_and_induced_coherence} we investigate the interference pattern and the visibility of the scheme from~\cite{Zou91} for arbitrary parametric gains of the two crystals similarly to~\cite{Belinsky92}.
Furthermore, we derive an expression for the coherence of the two interfering modes $1'$ and $2'$ as suggested by~\cite{Wiseman00}.

In \sref{sec_Regimes_of_induced_coherence} we use the previously obtained expressions to find the interference signal in the low-gain regime~\cite{Belinsky92}, an intermediate regime in analogy to~\cite{Shapiro15} and the high-gain regime.
The conclusion drawn in~\cite{Shapiro15} is that the visibility decreases when going from the low-gain regime to a high gain of crystal A.
The same holds if both crystals are strongly pumped.
However, in general there are two reasons for deterioration of the visibility in the interference pattern: (i) the intensities of two interfering waves become increasingly different, or (ii) the degree of coherence between these waves deteriorates.
Therefore, it is important to understand which one of two reasons (if not both) is responsible for deterioration of the visibility.

We discuss in detail both factors and determine in \sref{sec_Optimization} an optimal value for the visibility in the interference experiment as a function of the transmittance $T$.
We find that this optimal visibility does not deteriorate with increasing parametric gain of the crystal A but, on the contrary, improves.
Moreover, for high parametric gain and nonvanishing transmittance this optimum visibility attains values arbitrarily close to unity, as predicted in~\cite{Wiseman00} by the behavior of the coherence between the two signal waves $1'$ and $2'$.

Finally, we show in~\sref{sec_Signal_to_noise_ratio} that the signal-to-noise ratio improves for an increasing gain.
Unless a higher intensity would destroy the sample, there is no benefit for the signal-to-noise ratio to work in the low-gain regime, as already mentioned for the intermediate regime~\cite{Shapiro15}.
We then conclude in \sref{sec_Conclusions}.

For completeness, we describe in \ref{sec_Induced_coherence_in_the_Heisenberg_picture} each mode at every instance by bosonic creation and annihilation operators and calculate the photon number expectation values at the detectors D$_{1,2}$.

\section{Interference signal and induced coherence}
\label{sec_Interference_signal_and_induced_coherence}

To find a convenient treatment for the whole setup, we represent each mode $j =1,2,3,4$ as shown in \fref{fig_Setup} in terms of annihilation and creation operators $\hat{a}_{j}$ and $\hat{a}_{j}^{\dagger}$ so that the expectation value of the corresponding photon number operator gives the mean photon number of mode $j$.
A description of all annihilation operators at every instance of this experiment can be found in \ref{sec_Induced_coherence_in_the_Heisenberg_picture}.
We use these results to obtain the photon numbers $\hat{N}_{1,2}'' =  \hat{a}_{1,2}^{\prime\prime \dagger} \hat{a}_{1,2}''$ that are detected by the detectors D$_{1,2}$.

For a vacuum input, we find from \eref{eq_app_N12_expectation} the expression
\bg
\left\langle \hat N_{1,2}''\right\rangle= \frac{1}{2}\left(V_A+ V_B + V_A   V_BT \right) \left(1 \pm \mathcal{V} \cos 2\phi\right)\label{eq_N12}
\eg
for the expectation value in the two output ports.
Here, we defined the visibility
\bg
\mathcal{V}= 2 \frac{\sqrt{(1+V_A) V_A  V_B T  }}{V_A+ V_B +  V_A  V_B T}.
\label{eq_Visibility}
\eg
as well as the phase
\bg \label{eq_phi}
2 \phi \equiv \mathrm{arg}\left(u_A v_A v_B^*\right)
\eg
of the interference fringes, in complete agreement with the results of~\cite{Belinsky92}.
Note that $u_j$ and $v_j$ with $U_j-V_j\equiv |u_j|^2-|v_j|^2=1$ are the complex parameters of the Bogoliubov transformation from \eref{eq_bogu_A} and \eref{eq_bogu_B} describing the parametric process of crystal $j=$A,B with an undepleted pump.
Since the parameters $u_j$ and $v_j$ can be represented by hyperbolic functions, they scale exponentially with the parametric gain of the respective crystal.
Note that any phase shifter introduced in the idler mode between the crystals or in the signal modes shifts the phase of the interference pattern.

As noted in~\cite{Wiseman00}, the ultimate limit of the visibility is in fact given by the first-order coherence of the two arms before the final beam splitter, which can be quantified by the degree of coherence
\bg\label{eq_coherence}
\gamma_{12}\equiv\left| \left\langle \hat{a}_1^{\prime\dagger} \hat{a}_2^\prime\right\rangle\right|\Big/ \sqrt{\left\langle N_{1}'\right\rangle\left\langle N_{2}'\right\rangle}.
\eg
of the two  signal modes $1'$ and $2'$, where we defined the photon number operators $\hat{N}_j'=\hat{a}_j^{\prime\dagger }\hat{a}_j'$.
Note that this parameter describes the coherence induced between the two crystals, we will therefore refer to this quantity as \emph{induced coherence}.
With the help of~\eref{eq_trafo_A} and~\eref{eq_a2prime} we arrive at the expressions
\bg
\left| \left\langle \hat{a}_1^{\prime\dagger} \hat{a}_2^\prime\right\rangle\right|= \sqrt{  (1+ V_A) V_A V_B T }
\eg
as well as
\begin{equation}\label{e_N1prime}
\eqalign{
\left\langle \hat N_{1}'\right\rangle= V_A \cr
\left\langle \hat N_{2}'\right\rangle= (1+ T V_A) V_B,
}
\end{equation}
where we again assumed a vacuum input state in analogy to \ref{sec_Induced_coherence_in_the_Heisenberg_picture}.
With this insight, we find
\bg\label{eq_gamma}
\gamma_{12}=\sqrt{T\frac{1+V_A}{1+TV_A}}
\eg
for the induced coherence, in complete agreement with~\cite{Wiseman00}.
Note that it is independent of $V_B$, but changes depending on the transmittance $T$ and the gain in crystal A, parameterized by $V_A$.

\begin{figure}[htb] \centering
\includegraphics{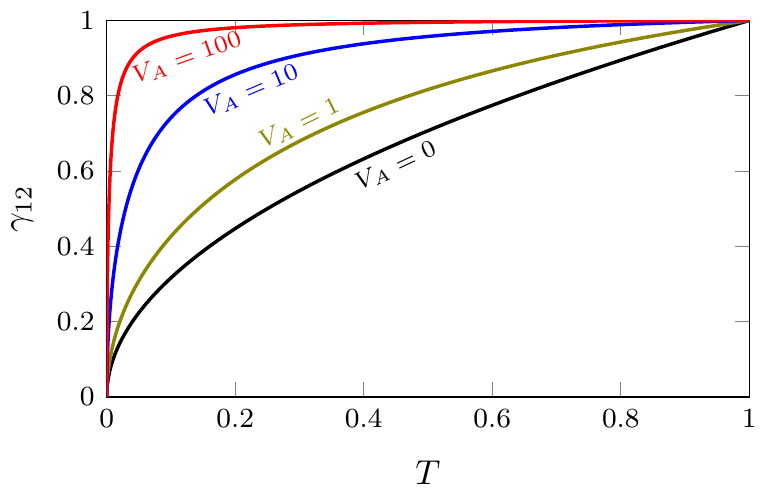}
    \caption{
    Dependence of the induced coherence $\gamma_{12}$ on the transmittance $T$ for different values of $V_A$.
    The coherence approaches unity for increasing gain $V_A T\gg 1$ and vanishes for $T=0$.
    The smallest coherence is obtained in the low-gain regime for $V_A\ll 1$ (black line).
	}
\label{fig_gamma_12}
\end{figure}

We plot the dependence of the coherence on $T$ in \fref{fig_gamma_12} for different values of $V_A$.
We see that the coherence increases with increasing gain and approaches the value of unity for $V_A T\gg 1$, which we show analytically in \sref{sec_Optimization}.
This fact has been already pointed out in~\cite{Wiseman00}, but we discuss in the following the connection to different cases in the literature, explain the diminishing visibility in the high-gain regime and find a way to optimize the visibility to its maximum given by the induced coherence.

\section{Regimes of induced coherence}
\label{sec_Regimes_of_induced_coherence}
We derived a general expression for the visibility of the interference signal and the induced coherence in the preceding section, and discuss now three special cases of interest:
(i)~The low-gain regime in which the interpretation from \sref{subsec_Setup_and_low-gain_interpretation} is valid and the experiments of~\cite{Zou91,Wang91,Barreto-Lemos14} were performed.
(ii)~The case of a high-gain source that is analyzed in~\cite{Shapiro15}, where the gain of crystal A is strong, but the one of crystal B is weak.
(iii)~We increase the gain in both crystals simultaneously and find a deterioration of the visibility for large transmittance in this high-gain regime.

\subsection{Low-gain regime}

The low-gain regime of spontaneous parametric down-conversion is also contained in our description.
We obtain this regime in the limit of vanishing gain, that is, for $V_A=V_B\ll 1$, where in fact induced emission is strongly suppressed~\cite{Zou91}.
In this case, we find from \eref{eq_Visibility} as well as \eref{eq_coherence}, in complete agreement with~\cite{Zou91,Wang91,Barreto-Lemos14}, that the visibility and induced coherence are related by
\bg\label{eq_V_lg}
\mathcal{V}^\mathrm{(lg)}= \sqrt{T}=\gamma_{12}^\mathrm{(lg)}.
\eg
We notice that in this limit the induced coherence from \eref{eq_gamma} coincides with the visibility, which means that the coherence induced in the modes $1'$ and $2'$ is the only limiting factor that reduces the visibility.
As we shall see later, this is not always the case.

\Eref{eq_V_lg} is plotted in \fref{fig_V_opt} as a black line.
This dependence on the transmittance has been verified multiple times experimentally and was seen as a quantum signature of the induced coherence.
However, in~\cite{Shapiro15} it is pointed out that in this limit the signal-to-noise ratio is very small.
This is due to the decreasing amplitude of the detected signal.
In fact, we find from \eref{eq_N12} the following relation for the expectation values
\bg
\left\langle \hat N_{1,2}''\right\rangle^\mathrm{(lg)}\cong V_A \left(1 \pm \sqrt{T} \cos 2\phi\right) \ll 1,
\eg
where we have to keep in mind that the number of photons produced in one crystal corresponds to $V_A \ll 1$ and is extremely small in the spontaneous regime.

\subsection{High-gain source}
\label{subsec_High-gain_source}

According to~\cite{Shapiro15}, increasing the gain in the source that induces the coherence, that is, crystal A, leads to a much better signal-to-noise ratio.
For the remainder of this article, we call this limit the regime of a \emph{high-gain source}.
In fact,~\cite{Shapiro15} refers to it as the \emph{classical} regime and contrasts it to the \emph{quantum} regime of spontaneous down-conversion discussed above.

We obtain the same results as~\cite{Shapiro15} in the limit $1 \ll V_A$ and $V_B \ll 1$, where we arrive by \eref{eq_Visibility} at the expression
\bg\label{eq_V_hgs}
\mathcal{V}^\mathrm{(hgs)}\cong 2 \sqrt{V_B T} \ll1
\eg
for the visibility with a high-gain source.
In this regime it is small and, in a naive comparison of \eref{eq_V_lg} with \eref{eq_V_hgs}, we in fact find $\mathcal{V}^\mathrm{(hgs)} \ll \mathcal{V}^\mathrm{(lg)}$.
However,~\cite{Shapiro15} points out that the signal-to-noise ratio is \emph{better} for a high gain source in comparison to the low-gain case.
This result can be seen directly from the relation
\bg
\left\langle \hat N_{1,2}''\right\rangle^\mathrm{(hgs)}\cong \frac{V_A}{2} \left(1 \pm 2 \sqrt{V_B T} \cos 2\phi\right) \gg 1,
\eg
where, when we identify $V_A \gg 1$ with the number of photons created in crystal A, we find a large signal with small visibility.
We  come back to this point when we discuss the signal-to-noise ratio in \sref{sec_Signal_to_noise_ratio}.

In contrast to the low-gain case, the visibility does not correspond to the coherence between the two arms, and, in particular, is significantly smaller than the coherence function (for $T\neq 0$).

\subsection{High-gain regime}
\label{subsec_High-gain_regime}

\begin{figure}[htb] \centering
\includegraphics{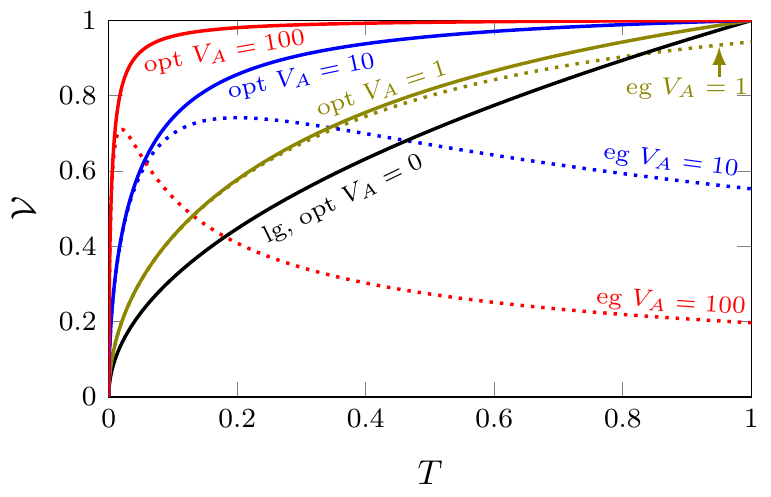}
    \caption{
    Visibility of the interference signals for different regimes and different gain $V_A$.
    The dotted lines show the visibility $\mathcal{V}^\mathrm{(eg)}$, given by \eref{eq_V_ep}, for equal pumping ($V_A=V_B$), the solid lines show the optimal visibility $\mathcal{V}^\mathrm{(opt)}$, defined in \eref{eq_V_opt}.
    Note that the optimal visibility corresponds to the induced coherence, that is, $\mathcal{V}^\mathrm{(opt)}=\gamma_{12}$, so that the solid lines are exactly the same as in \fref{fig_gamma_12}.
    The low-gain visibility  $\mathcal{V}^\mathrm{(lg)}$ is obtained for $V_A=0$ (black line).
	}
\label{fig_V_opt}
\end{figure}

In this section we investigate the influence of increased gain in both crystals.
For that, we assume equal gain in both crystals, that is, $V_A=V_B$, and find from \eref{eq_Visibility} that the visibility is given by
\bg\label{eq_V_ep}
\mathcal{V}^\mathrm{(eg)}= 2\frac{\sqrt{(1+V_A) T} }{2+ V_A T}.
\eg
We plot this expression for the visibility with dotted lines for different values of $V_A$ in \fref{fig_V_opt}.

For low transmittance and high gain we have a behavior that is close to the induced coherence $\gamma_{12}$ (the respective solid lines) that corresponds to the optimal visibility (see \sref{sec_Optimization}), exceeding the low-gain visibility (black line).
However, for increasing gain and higher transmittance, the visibility drops significantly below the low-gain result.
There is an intuitive explanation for this fact: due to seeding by crystal A, crystal B produces more photons than the first one.
Hence, both arms become unevenly populated and the visibility drops for increasing gain.
But if the transmittance is sufficiently small, the seeding effect is suppressed (since most photons from crystal A are not transmitted) and both arms have roughly the same intensity.
This explains why the visibility follows the coherence for a small transmittance and diminishes for a large one.

\section{Optimization}
\label{sec_Optimization}

As pointed out in the previous section, the visibility is limited not only by the induced coherence, but also by the intensity difference in the two modes $1'$ and $2'$ before the second beam splitter.
With the help of \eref{e_N1prime} we find that for
\bg\label{eq_opt_V_B}
V_B = \frac{V_A}{1+  V_A T}
\eg
the photon-number difference $\langle \hat N_{1}'\rangle-\langle \hat N_{2}'\rangle$ vanishes.
By adjusting the pump of crystal B (varying $V_B$) according to the strength of the seed (which is determined by $T$ and $V_A$), the photon number produced by crystal B changes in such a way that it corresponds to the photon number in the other arm of the interferometer.
In fact, with this choice of $V_B$ we arrive with \eref{eq_Visibility} at the visibility
\bg\label{eq_V_opt}
\mathcal{V}^\mathrm{(opt)}=\sqrt{T\frac{1+V_A}{1+TV_A}}= \gamma_{12},
\eg
where we identified the optimized visibly with the induced coherence from \eref{eq_gamma}.

We depict $\mathcal{V}^\mathrm{(opt)}$ (and by that the induced coherence $\gamma_{12}$) for different gain values in \fref{fig_V_opt} (solid lines) and see that it always exceeds the low-gain limit of spontaneous parametric down-conversion (black line) as well as the visibility for equal pumping, that is, $\mathcal{V}^\mathrm{(ep)}\leq \mathcal{V}^\mathrm{(opt)}=\gamma_{12}$.
In fact, we achieve almost perfect visibility in the high-gain regime, which is apparent when we expand \eref{eq_V_opt} for $TV_A\gg 1$ giving rise to
\bg\label{eq_hg_opt_vis}
\mathcal{V}^\mathrm{(opt)}\cong 1 - \frac{1-T}{2}\frac{1}{TV_A}+ \mathcal{O}\left(\frac{1}{T^2V_A^2}\right).
\eg
It is remarkable to note that in the low-gain regime a low transmittance limits the visibility, whereas the optimized version in the high-gain regime approaches unity even for small $T$.
This fact can be used to enhance the distinction between two slightly different small values of $T$ and perform high-contrast imaging of objects with low transmittance.

However, we obtain not only the optimal visibility close to unity, but also find an increased signal.
Using again the adjusted $V_B$ from \eref{eq_opt_V_B} in \eref{eq_N12}, we arrive at
\bg
\left\langle \hat N_{1,2}''\right\rangle^\mathrm{(opt)}= V_A \left(1 \pm \sqrt{T\frac{1+V_A}{1+V_A T}} \cos 2\phi\right).
\eg
Therefore, the overall amplitude of the interference signal also scales with the number of photons produced by crystal A.

To investigate how the transmittance $T$ affects the coherence of the two crystals we compare it to a more general setup.
For that, we introduce an additional beam splitter with transmittance $T_2$ in the mode $2'$ after crystal B.
By setting either $T$ or $T_2$ to unity, we can switch between two different situations: (i) a non-unity transmittance after crystal B to attenuate the population of mode $2'$ and (ii) a transmittance between the crystal that controls the induced coherence.

Performing exactly the same analysis with an additional beam splitter of transmittance $T_2$ in the output signal mode of crystal B leads to the same expectation value from \eref{eq_N12}, where only $V_B$ is replaced by $T_2 V_B$.
Hence, the condition for optimal visibility reads
\bg \label{e_T2Vb_opt}
T_2 V_B = \frac{V_A}{1+  V_A T}
\eg
in analogy to \eref{eq_opt_V_B} and can be easily interpreted: equal intensities in both arms cannot only be found by adjusting the gain of crystal B, but also by modulating the output of crystal B with the help of the transmittance $T_2$.
Hence, using the additional beam splitter only introduces a second parameter to control this photon number.

\section{Signal-to-noise ratio}
\label{sec_Signal_to_noise_ratio}

Considering that the effect of induced coherence without induced emission was used in~\cite{Barreto-Lemos14} as an imaging technique, not only the visibility is essential, but also the signal-to-noise ratio.
In fact, it was pointed out in~\cite{Shapiro15} that for a high-gain source the signal-to-noise ratio is improved in comparison to the low-gain regime.
In this section we extend this discussion to the optimized scheme as well as the high-gain regime and show that in both cases the signal-to-noise ratio is even higher.

\subsection{General expression}

Since in the imaging experiment of~\cite{Barreto-Lemos14} the image was obtained from the difference of the two output signals $\hat{N}_-\equiv \hat{N}_{1}''-\hat{N}_2''$, we restrict our treatment to the same quantity.
The signal-to-noise ratio can be defined as the square of the expectation value divided by the variance, that is,
\bg
\mathrm{SNR}_- = \left\langle\hat{N}_-\right\rangle^2\bigg/\left\langle\Delta \hat{N}_-^2\right\rangle.
\eg
From \eref{eq_N12} it is easy to see that the expectation value of the photon number difference takes the form
\bg\label{eq_N_-}
\left\langle\hat{N}_-\right\rangle = 2 \sqrt{(1+V_A)V_A V_B T}\cos 2 \phi.
\eg
With the operators obtained in \ref{sec_Induced_coherence_in_the_Heisenberg_picture} we find after a straightforward, but cumbersome, calculation that
\bg \label{eq_SNR_general}
\mathrm{SNR}_-=\frac{\left\langle\hat{N}_-\right\rangle^2}{\left\langle\hat{N}_-\right\rangle^2+ V_A+V_B+V_A V_B (2- T)}
\eg
for the vacuum expectation value.
In the following we investigate this expression in the different regimes.

Note that the equations above as well as in the remainder of this section describe the signal-to-noise ratio for a single mode.
However, our treatment can be generalized to the measurement of multiple modes, for example to many pump pulses in analogy to~\cite{Shapiro15}.
In fact, we can define the normalized multi-pulse photon-number difference for $p$ pulses (where the previously obtained photon numbers are now labelled by an additional index for the $j$th pulse).
Its expectation value corresponds to the one obtained for a single pulse, namely 
\bg
\left\langle \hat{N}_-^{(p)} \right \rangle = \frac{1}{p} \sum\limits_{j=1}^p \left\langle\hat{N}_{1}''(j)-\hat{N}_{2}''(j) \right\rangle = \left\langle \hat{N}_- \right\rangle.
\eg
Note that we assumed equal gain for all pulses and thus the summation can be performed trivially.
We see that the normalized difference of the two output signals remains unchanged and equal to Eq.~\eref{eq_N_-}.
In the same way, multiple pulses increase the number of photons produced if the signals of the two detectors are not subtracted.
However, the visibility per pulse---by definition normalized to unity---is independent of the number of pulses $p$.

In contrast to this result, we find for the multi-pulse variance
\bg
\left\langle \Delta \hat{N}_-^{(p)\; 2} \right\rangle =\frac{1}{p}\left\langle \Delta \hat{N}_-^2 \right\rangle
\eg
and therefore
\bg
\mathrm{SNR}_-^{(p)}= p~\mathrm{SNR}_-.
\eg
Hence, the signal-to-noise ratio is increased by (and proportional to) the number of pulses $p$.
In the following we discuss only the signal-to-noise ratio of a single pulse.
We will see that it does not exceed unity and can take very small values.
However, we emphasize that our results can be easily generalized to a multi-pulse treatment by multiplying them with the number of pulses.
This way, values larger than unity can be experimentally achieved.

\subsection{Low-gain regime}
\label{subsec_SNR_Low-gain_regime}

\begin{figure}[htb] \centering
\includegraphics{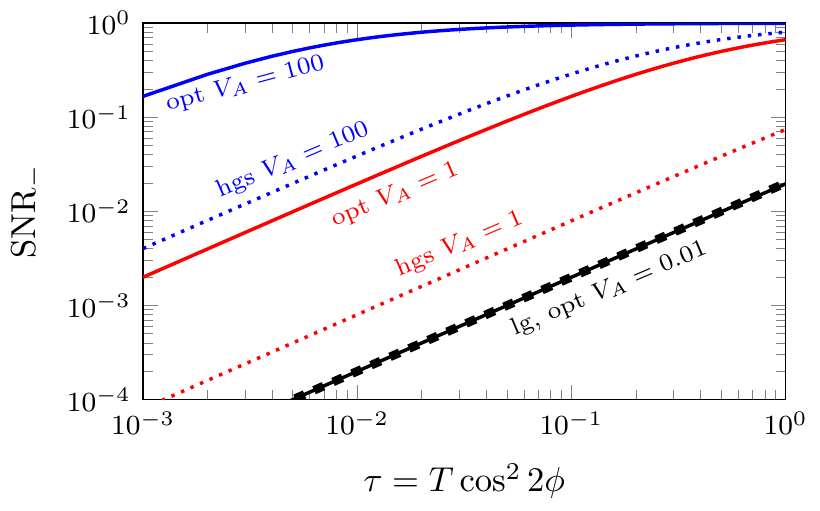}
    \caption{
    Signal-to-noise ratio of the difference of photon numbers in the two exit ports.
    The dependence on $\tau = T \cos^22\phi$ reflects the fact that this setup can be used for phase and absorption imaging.
    We see that the low-gain approximation (lg, dashed thick line) scales with the parameter $V_A$.
    The ratio increases for an increasing gain in crystal A (hgs, dotted thin lines), where we used the low-gain value of $V_B=0.01$.
    It is even better if the gain is optimized and the crystals are operated in the high-gain regime, as we see for different values of $V_A$ (opt, solid lines).
	}
\label{fig_SNR}
\end{figure}

When we take the low-gain limit in both crystals ($V_A=V_B \ll 1$) we arrive with the help of \eref{eq_N_-} and \eref{eq_SNR_general} at the expression
\bg\label{eq_SNR_lg}
\mathrm{SNR}_-^\mathrm{(lg)}\cong 2 T V_A \cos^2 2\phi \ll 1
\eg
for the signal-to-noise ratio in accordance with~\cite{Shapiro15}.
Because the overall amplitude scales with the number of photons generated in one crystal, the signal-to-noise ratio in the low-gain limit is very small.
We see this effect when we plot $\mathrm{SNR}_-^\mathrm{(lg)}$ as a function of $\tau =T \cos^2 2 \phi$ in \fref{fig_SNR} (dashed thick line) for $V_A=0.01$.
The fact that $\tau$ depends on two parameters $\phi$ and $T$ accounts for the possible application in phase and absorption imaging.

\subsection{High-gain source}

In contrast to the low-gain regime, a high-gain source gives
\bg
\mathrm{SNR}_-^\mathrm{(hgs)}\cong \frac{4 (1+V_A) V_B T\cos^2 2\phi}{1+4 (1+V_A) V_B T\cos^2 2\phi}
\eg
in the limit of $V_B \ll 1$ and $V_A\gg 1$ using \eref{eq_N_-} and \eref{eq_SNR_general}.

This result, as well as \eref{eq_SNR_lg},  were already obtained in~\cite{Shapiro15}, where it was demonstrated that $\mathrm{SNR}_-^\mathrm{(hgs)}$ is larger than $\mathrm{SNR}_-^\mathrm{(lg)}$.
We show this effect in \fref{fig_SNR}, where we plot $\mathrm{SNR}_-^\mathrm{(hgs)}$ for different gain values and depending on $\tau= T \cos^22\phi$.
In fact, the low-gain result for $V_A=0.01$ (dashed thick line) is smaller in comparison to the high-gain source (thin dotted lines) for  $V_B=0.01$ and different $V_A$.
Thus, increasing the gain of crystal A is beneficial for imaging, as already implied in \sref{subsec_High-gain_source} and pointed out by~\cite{Shapiro15}.

\subsection{Optimized gain}

Since our treatment is general enough to include the high-gain regime for both crystals, we demonstrate that this limit is even more beneficial from the point of view of the signal-to-noise ratio.
For that, we first discuss the case of an optimized gain following the results from \sref{sec_Optimization}.

When we choose $V_B$ according to \eref{eq_opt_V_B} so that the visibility is optimal, we arrive with \eref{eq_N_-} and \eref{eq_SNR_general} at the expression
\bg\label{eq_SNR-opt}
\mathrm{SNR}_-^\mathrm{(opt)}= \frac{2 V_A T\cos^2 2\phi}{1+2 V_A T\cos^2 2\phi},
\eg
which we plot as function of $\tau= T \cos^22\phi$ in \fref{fig_SNR} for increasing values of $V_A$.
We see that it approaches unity for $V_A \tau \gg 1 $ and that $\mathrm{SNR}_-^\mathrm{(opt)}$ (solid lines) is even larger than $\mathrm{SNR}_-^{\mathrm{(hgs)}}$ (thin dotted lines).
Note further that this equation reduces to $\mathrm{SNR}_-^\mathrm{(lg)}$ for $V_A \ll 1$.
In fact, for small parametric gain we obtain the low-gain limit from \sref{subsec_SNR_Low-gain_regime} (the thick dashed and the black solid line are on top of each other).

\subsection{High-gain regime}

One could imagine that the signal-to-noise ratio is better for an optimized visibility compared to the high-gain regime with an equal pumping in both crystals, where the visibility deteriorates.
However, we shall show in the following that this reasoning is misleading.

For that, we assume equal gain in both crystals, set $V_A=V_B$ in \eref{eq_N_-} and \eref{eq_SNR_general}, and arrive at
\bg\label{eq_SNR_eg}
\mathrm{SNR}_-^\mathrm{(eg)}= \frac{2 V_A T \cos^2 2 \phi}{1+ 2 V_A T \cos^2 2 \phi - T V_A/(2+2V_A) }.
\eg
When we now consider the ratio
\bg\label{eq_SNR_ratio}
\frac{\mathrm{SNR}_-^\mathrm{(opt)}}{\mathrm{SNR}_-^\mathrm{(eg)}}=1- \frac{T V_A}{2(1+V_A)(1+2V_A T \cos^22\phi)}\leq 1,
\eg
we see that even though the visibility is optimized, $\mathrm{SNR}_-^\mathrm{(opt)}$ is always \emph{smaller} when compared to an equal gain in both crystals for $T\neq 0$.
This can be understood by the fact that the overall signal is smaller if crystal B is adjusted for optimal visibility.
However, we see from \eref{eq_SNR_ratio} that this difference vanishes in the high-gain regime ($ V_A T \cos^2 2\phi\gg 1$).
Since the behavior is almost exactly the one of \eref{eq_SNR-opt}, we refrain from plotting additional curves but emphasize that the respective lines would follow closely the solid ones in \fref{fig_SNR}.

\section{Conclusions}
\label{sec_Conclusions}

In this article we have shown that the effect of induced coherence persists in different regimes of parametric gain for two crystals.
Whereas in the low-gain regime of spontaneous parametric-down conversion with suppressed induced emission the induced coherence is limited by the transmittance in the idler mode between two crystals, and, therefore, by the distinguishability of the two sources, it approaches unity in the high-gain regime~\cite{Wiseman00}.
However, due to intensity difference of two signal waves before the final beam spitter, the visibility deteriorates with increasing gain.
This effect can be compensated by adjusting the gain in the second crystal.

Induced coherence can be used to perform imaging~\cite{Barreto-Lemos14}, but the signal-to-noise ratio is small in the low-gain regime~\cite{Shapiro15}.
It can be improved by operating crystal A with high parametric gain and even more when both crystals are operated in this regime.

Our results can be generalized to the case where a nonvanishing classical field seeds the input of one of the crystals, which we plan to address in the future.
Let us note that we considered in this article a classical pump neglecting depletion.
Using such a strong pump field leads necessarily to higher-order photon creation in both crystals.
Therefore, a seeding effect of crystal B by an idler photon from crystal A can never be excluded, even in the low-gain regime, where these events are extremely rare.

Our results represent a comprehensive treatment of the effect of induced coherence with a classical pump, explain different regimes discussed in the literature~\cite{Belinsky92,Wiseman00,Shapiro15} and show the potential for possible applications for imaging in the high-gain regime. In this regime, we observe a better signal-to-noise ratio and can optimize the visibility such that the contrast for small transmittance is enhanced.

\ack
We thank M. Lahiri for fruitful discussions.
MIK acknowledges  partial financial support by the European Union's Horizon 2020 research and innovation programme under grant agreement No 665148 (QCUMbER).
He also thanks the Max Planck Centre for Extreme and Quantum Photonics, University of Ottawa, where part of this work was accomplished, for its hospitality during his stay.
SL, RF, and RWB gratefully acknowledge support by the Canada Excellence Research Chairs program and the Natural Sciences and Engineering Research Council of Canada (NSERC).
RF acknowledges the support of the Banting postdoctoral fellowship of the NSERC and SL the financial support from Le Fonds de Recherche du Qu\'{e}bec Nature et Technologies (FRQNT).
EG is grateful to the Friedrich-Alexander-Universit\"at Erlangen-N\"urnberg for an Eugen Lommel stipend.
\appendix

\section{Heisenberg picture}
\label{sec_Induced_coherence_in_the_Heisenberg_picture}

We shall consider the scheme of \fref{fig_Setup} in the Heisenberg representation and describe each mode $j=1,2,3,4$ of the nonlinear interferometer in terms of annihilation and creation operators $\hat{a}_{j}$ and $\hat{a}_{j}^{\dagger}$, obeying standard single-mode commutations relations, $[\hat{a}_{j},\hat{a}_{k}^{\dag}]=\delta_{j,k}$, and normalized so that
$\langle \hat{N}_{j} \rangle=\langle \hat{a}_{j}^{\dagger} \hat{a}_{j} \rangle$ gives the mean photon number in the corresponding mode $j$.
This description is in analogy to~\cite{Belinsky92,Wiseman00}.

We start by noting that the transformation
\begin{equation}\label{eq_trafo_A}
\eqalign{
\hat{a}_1' = u_A \hat{a}_1 + v_A \hat{a}_3^\dagger \cr
\hat{a}_3' = u_A \hat{a}_3 + v_A \hat{a}_1^\dagger
}
\end{equation}
describes the photon annihilation operators in the output modes $1'$ and $3'$ of crystal A for an undepleted classical pump.
Here, $\hat{a}_1$ and $\hat{a}_3$ describe the photon annihilation operators for the two input modes according to \fref{fig_Setup}.
Note that this transformation corresponds to a unitary Bogoliubov transformation with
\bg \label{eq_bogu_A}
1=|u_A|^2-|v_A|^2\equiv U_A - V_A.
\eg
Furthermore, $u_A$ and $v_A$ are in general complex and $U_A$ and $V_A$ can be represented by respective hyperbolic functions.

The transmittance in the idler modes between the crystals A and B leads to the transformation
\begin{equation}
\eqalign{
\hat{a}_3''= t \hat{a}_3' + r \hat{a}_4 \cr
\hat{a}_4'= t \hat{a}_4 - r \hat{a}_3'
}
\end{equation}
for the annihilation operators in the output modes $3''$ and $4'$ of the beam splitter S$_1$.
Here, we assumed $t,r \in \mathbb{R}$, which fulfil the relation
\bg\label{eq_transmit}
1= t^2 + r^2 \equiv T + R,
\eg
where $T$ and $R$ denote the transmittance and reflectivity of the intensity.
The annihilation operators $\hat{a}_3'$ and $\hat{a}_4$ represent the input modes of the beam splitter.
Since $\hat{a}_3'$ is the output of crystal A, we find from \eref{eq_trafo_A} the expression
\bg\label{eq_a3doubleprime}
\hat{a}_3''= t v_A \hat{a}_1^\dagger + t u_A \hat{a}_3 + r \hat{a}_4 .
\eg

The transformation
\begin{equation}
\label{eq_trafo_B}
\eqalign{
\hat{a}_2' = u_B \hat{a}_2 + v_B \hat{a}_3^{\prime\prime\dagger}\cr
\hat{a}_3''' = u_B \hat{a}_3'' + v_B \hat{a}_2^\dagger
}
\end{equation}
describes the action of crystal B and fulfils, in analogy to \eref{eq_bogu_A}, the relation
\bg\label{eq_bogu_B}
1=|u_B|^2-|v_B|^2\equiv U_B - V_B.
\eg
Since crystal B is seeded by one of the output modes of beam splitter S$_1$, we find with the help of \eref{eq_a3doubleprime} the expression
\bg\label{eq_a2prime}
\hat{a}_2' = t v_A^* v_B \hat{a}_1 + u_B \hat{a}_2 + t u_A^* v_B \hat{a}_3^\dagger + r v_B \hat{a}_4^\dagger,
\eg
where $*$ denotes the complex conjugate.

So far, we have obtained the expression for the annihilation operators in the modes $1'$ and $2'$, which only depend on the input modes $1,2,3$ and $4$.
These two modes interfere at the final beam splitter S$_2$, which we assume to be a 50:50 beam splitter, and thus fulfil the relation
\begin{equation}
\eqalign{\hat{a}_1''= \left(\hat{a}_1' + \hat{a}'_2\right)/\sqrt{2}\cr
\hat{a}_2''= \left(\hat{a}_1' - \hat{a}'_2\right)/\sqrt{2}}
\end{equation}
describing the two output modes $1''$ and $2''$.
With \eref{eq_trafo_A} and \eref{eq_a2prime} we therefore arrive at
\begin{equation}
\eqalign{
\hat{a}_{1,2}''=\frac{1}{\sqrt{2}}\big[& \left( t v_A^* v_B \pm u_A \right)\hat{a}_1+ u_B \hat{a}_2 \cr
&+ \left(t u_A^* v_B \pm v_A \right) \hat{a}_3^\dagger + r v_B \hat{a}_4^\dagger\big]
}
\end{equation}
for the annihilation operators before detection.

With these results we are able to calculate the expectation values of the photon numbers $\hat{N}_{1,2}''=\hat{a}_{1,2}^{\prime \prime \dagger} \hat{a}_{1,2}''$ in the two exit ports and arrive at
\begin{equation}
\label{eq_app_N12}
\eqalign{\hat{N}_{1,2}''= \frac{1}{2}\Big[& \left| t v_A^* v_B \pm u_A \right|^2 \hat{N}_1 + U_B \hat{N}_2+ R V_B \left(\hat{N}_4+1\right)\\
& +  \left|t u_A^* v_B \pm v_A \right|^2 \left(\hat{N}_3+1\right) \\
& + \mathrm{~combinations~of~input~modes}\Big].
}
\end{equation}
Since the crystals is not seeded, we assume a vacuum input with $\langle\hat{a}_j^\dagger \hat{a}_k \rangle=0$ for all combinations $j,k=1,2,3,4$.
With this insight and the definition
\bg
t u_A v_A  v_B^* \equiv \sqrt{T  U_A V_A V_B}\exp(\I 2 \phi)
\eg
we use~\eref{eq_bogu_A},~\eref{eq_transmit}, as well as~\eref{eq_bogu_B} to find from \eref{eq_app_N12} the expectation values
\begin{equation}
\label{eq_app_N12_expectation}
\eqalign{
\left\langle \hat{N}_{1,2}''\right\rangle= \frac{1}{2}\Big[&  (V_A+V_B+T V_A V_B)\cr
& \pm 2 \sqrt{T  U_A V_A V_B} \cos(2\phi) \Big]
}
\end{equation}
for the photon number at the detector D$_1$ and D$_2$.

\section*{\refname}

\bibliographystyle{unsrt}

\bibliography{bibliography}

\end{document}